# Complex lattice and charge inhomogeneity favoring quantum coherence in high temperature superconductors[1]

Antonio Bianconi[1-5]

[1] RICMASS, Rome International Center for Materials Science Superstripes, via dei Sabelli 119A, 00185 Rome, Italy.
[2] Institute of Crystallography, CNR, via Salaria Km 29.300, Monterotondo Roma, I-00015, Italy
[3] Solid State and Nanosystems Physics, National Research Nuclear University "MEPhI" (Moscow Engineering Physics Institute), Moscow, Ru
[4] Department of high temperature superconductivity and nanostructures, Solid State [4]Physics division, P.N. Lebedev Physical Institute of the Russian Academy of Sciences, Moscow, Ru
[5] Consorzio Interuniversitario INSTM, Udr Rome, Italy

Abstract

The presence of two components in the electron fluid of high temperature superconductors and the complex charge and lattice inhomogeneity have been the hot topics of the international conference of the superstripes series, *Superstripes 2015,* held in Ischia in 2015. The debate on the mechanisms for reaching room temperature superconductors has been boosted by the discovery of superconductivity with the highest critical temperature in pressurized sulfur hydride. Different complex electronic and structural landscapes showing up in superconductors, which resist to the decoherence effects of high temperature, have been discussed. While low temperature superconductors described by the BCS approximation are made of a single condensate in the weak coupling the high temperature superconductors are made of coexisting multiple condensates (in different spots of the k-space and the real space) some in the BCS-BEC crossover regime and others in the BCS regime. The role of "shape resonance" in the exchange interaction between these different condensates, like "the Fano-Feshbach resonance" in ultracold gasses, is emerging as a key term for high temperature superconductivity.

**keywords:** Superstripes, Multi-condensates superconductivity; Nanoscale textures; Lifshitz transitions; BCS-BEC crossover; Shape resonances; Fano resonances.

---

[1] *Preface for the special issue of* Journal of Superconductivity and Novel Magnetism *for the International Conference "SUPERSTRIPES 2015" held in Ischia (Italy) June 13-18, 2015*

The major new scientific achievement presented at the Superstripes 2015 conference, held in June 2015 in the splendid landscape of the Ischia Island, has been the report of Meissner effect and the record for the highest superconductivity critical temperature, $T_c$= 203 K, in pressurized sulfur hydride [1] which has attracted high interest [2]. The high temperature superconductivity in this metallic hydride emerges when the pressure induces the disproportionation $3(H_2S) \rightarrow 2(H_3S) + S$ with phase separation between a first cubic $H_3S$ phase coexisting with a second S phase. Superconductivity onset occurs where the pressure tunes the chemical potential of metallic $H_3S$ around a Lifshitz transition for the appearing of new Fermi surface spots at the gamma point of the Brillouin zone and at the predicted structural phase transition from R3m to Im-3m crystal symmetry. The maximum critical temperature occurs at a second electronic topological Lifshitz transition Fermi for a neck disrupting Fermi surface topology where a new two-dimensional Fermi surface spot appears [3,4].

The different types of superconducting phases in all known high temperature superconductors seem to have in common the formation of an exotic superconducting phase made of multiple condensates which has been described by the BPV (Bianconi-Perali-Valletta) theory [5-9], where the chemical potential is tuned near Lifshitz transitions [10,11] by external or internal chemical pressure. Near a Lifshitz transition the Migdal approximation breaks down in the new appearing Fermi surface spot and polaronic charge carriers are formed in the new appearing Fermi surface. These carriers form a condensate in the BCS-BEC crossover which reminds the bipolaronic condensate proposed by K.A. Muller [12]. In the formation of the superconducting condensates [13-15] the exchange-like interaction between bipolarons in the appearing small Fermi surface spots and pairs of electrons in other large Fermi surface spots can give shape resonances in the superconducting gaps which provide a contact pairing interaction to boost superconductivity, like the Fano-Feshbach resonances [16-24] which have been well studied in ultracold gases.

The spatial complexity emerges in these systems since by tuning the chemical potential near a Lifshitz transition the electronic system made by strongly interacting particles is predicted to undergo phase separation [25,26] which can be frustrated by a long range Coulomb interaction giving a multiscale phase separation going from the nanoscale to the micron scale spanning the mesoscale spatial range between the atomic and macroscopic space. In fact phase separation has been observed in cuprates and related materials [27-32] and it has been related with exotic theories of high temperature superconductivity [33-43]

The complex electronic and structural landscape of the strongly correlated electronic structure of the quasi 2D copper oxide plane in cuprate superconductors, was first unveiled by XANES spectroscopy using synchrotron radiation [44-48]. It has provided first, the evidence of the orbital character of the itinerant holes (in the oxygen 2p orbital) [49] and second, the coexistence of two electronic states at the Fermi level: polarons [50] (where the ratio between the pairing interaction and the local Fermi energy is close to one) around the antinodal point and free quasi-particles around the nodal point of the Brillouin zone. Ordering of polarons gives the *superstripes* scenario [51] where inhomogeneous charge density wave (CDW) order coexists and competes with superconducting order [52]. A similar scenario occurs in iron-based superconductor where polarons in a shallow Fe($3d_{xz}$,$3d_{yz}$) orbital band, forming a very small Fermi surface, coexist with free quasi-particles in multiple large Fermi surfaces [53]. The two electronic components segregate forming a nanoscale phase separation, extending in the mesoscale, forming non Euclidean geometries for the interstitial superconducting phase [52]. Complexity and phase separation scenarios in complex oxides and functional strongly correlated systems have recently attracted a wide interest [53-63]. Iron-based superconductors have provided a key laboratory showing exotic superconducting phases characterized by inhomogeneity, nematic phases and multiple gaps [64-69]. The physics of the BCS-BEC crossover in quantum condensates in particular real space or k space spots widely studied in ultracold gases is now emerging as a generic feature in multiple condensates superconductors [70-76]. Understanding the emergence of novel spin-orbital phases [77] is now recognized to be a key ingredient for understanding high temperature superconductors and the advances in this field are now pushing the developments of spintronics and new quantum electronics [78-83]. The isotope coefficient in the superconducting critical temperature is predicted to be independent on the variation of the chemical potential in the BCS single band theory. Therefore the experimentally observed changes of the isotope coefficient with variable doping in cuprates [84,85] and with variable pressure in sulfur hydride [2], have been object of discussion as the *smoking gun* for exotic superconducting phases made of multiple condensates where some of them are close to a band edge, and more works in this field are expected to come. Finally the superstripes conference series is becoming the major international forum on the emergence of quantum mechanics in macroscopic properties of complex matter, where the scientific advances in atomic ultracold condensates, and the science of complex materials meet opening new perspective for understanding quantum biology and promoting quantum electronics.